\documentclass[12pt]{article}

\usepackage{amsmath,amssymb}
\usepackage{graphicx}
\usepackage{dcolumn}
\usepackage{cite}
\oddsidemargin  - 5mm
\evensidemargin - 5mm

\newcommand{\be}{\begin{equation}}
\newcommand{\ee}{\end{equation}}
\newcommand{\bea}{\begin{eqnarray}}
\newcommand{\eea}{\end{eqnarray}}

\newcommand{\sot}{SO(3, \mathbb{R})}

\begin{document}

\title{\bf Light-cone SU(2) Yang-Mills theory and conformal mechanics}

\author{
V. Gerdt~$^a$,
A. Khvedelidze~$^{a,\ b,\ c}$, and
D. Mladenov~$^d$\\[0.7cm]
$^a$
{\it Joint Institute for Nuclear Research, Dubna 141980, Russia.}\\[0.3cm]
$^b$
{\it A. Razmadze Mathematical Institute, Tbilisi GE-0193, Georgia.}\\[0.3cm]
$^c$
{\it University of Plymouth, Plymouth PL4 8AA, United Kingdom.}\\[0.3cm]
$^d$
{\it DESY, Hamburg 22607, Germany.}
}

\date{\empty}
\maketitle

\begin{abstract}
We examine the mechanical matrix model that can be derived from
the $SU(2)$ Yang-Mills light-cone field theory by restricting the
gauge fields to depend on the light-cone time alone. 
We use Dirac's generalized Hamiltonian approach. 
In contrast to its
well-known instant-time counterpart the light-cone version of
$SU(2)$ Yang-Mills mechanics has in addition to the constraints,
generating the $SU(2)$ gauge transformations, the new first and
second class constraints also. 
On account of all of these
constraints a complete reduction in number of the degrees of
freedom is performed. 
It is argued that the classical evolution of the 
unconstrained degrees of freedom is equivalent to a free
one-dimensional particle dynamics. 
Considering the complex
solutions to the second class constraints we show at this time
that the unconstrained Hamiltonian system represents the
well-known model of conformal mechanics with a ``strength'' of the
inverse square interaction determined by the value of the gauge
field spin.
\end{abstract}

\newpage


\section{Introduction}


Nowadays the correspondence between gauge theories in various
dimensions and integrable systems has become a subject of
intensive study. 
After the pioneering work by Seiberg and Witten
\cite{SeibergWitten}, demonstrating that $N=2$ supersymmetric
Yang-Mills theory in four dimensions is exactly solvable in the
low-energy limit, considerable progress in the understanding of
these relations has been marked. 
In scope of the correspondence to
the underlying integrable systems several properties of the
Seiberg-Witten theory have been investigated using the approach
proposed in \cite{GKMMM}. 
In particular, it was shown that the
low-energy effective action can be described in terms of different
one-dimensional integrable many-body systems ranging from the
classical Toda-chain model in the case of supersymmetric
Yang-Mills theory without matter, to elliptic Calogero-Moser model
when adjoint matter is added and to classical spin {\bf XXX} chain
for theory with fundamental matter included (for comprehensive
reviews of these studies see e.g.
\cite{GorskyMironov,Marshakov,DHokerPhong:Lectures}).

At the same time similar relationships have been observed also for
non-super\-sym\-metric gauge theories. 
It was recognized that the
{\bf XXX} Heisenberg spin chains are related to other physically
interesting limits in QCD. 
Namely, an equivalence was found
between the Hamiltonian describing the Regge asymptotic behavior
of hadron-hadron scattering amplitudes in QCD and the Hamiltonian
of the $SL(2, \mathbb{C})$ {\bf XXX} Heisenberg magnet
\cite{QCD-Regge}. 
Furthermore it turns out that the logarithmic
evolution of the composite operators in QCD on the light-cone is
similar to the dynamics of $SL(2, \mathbb{R})$ {\bf XXX}
Heisenberg spin chain \cite{QCD-LC}. 
Based on this hidden
integrability of the effective theories of QCD a kind of
stringy/brane picture was developed recently \cite{Gorsky-HEQCD}.

There is also a physically very important regime when
finite-dimensional system arises in the context of gauge field
theory. 
The long-wavelength approximation in the dynamics of gauge
fields effectively leads to the so-called dimensional $(1 + 0)$
reduction of the field theory and at first has been intensively
studied for the non-supersymmetric Yang-Mills theory, both from
physical as well as from a purely mathematical point of view (see
e.g \cite{MatinyanSavvidyTer-Arutyunyan-Savvidy}-\cite{KM} and
references therein). 
In the middle of 1980's analogous
supersymmetric mechanical models with more than four
supersymmetries were constructed from the corresponding super
Yang-Mills theory
\cite{ClaudsonHalpern,Flume,BaakeReinickeRittenberg}. In
particular, in \cite{ClaudsonHalpern} the maximally supersymmetric
$N=16$ gauge mechanics was considered. 
The recent renewed interest
in the supersymmetric version of Yang-Mills mechanics is motivated
by the observation that the Hamiltonian of $D=1$ $SU(n)$ super
Yang-Mills theory in the large $n$ limit describes the dynamics of
$D=11$ supermembrane \cite{deWitHoppeNicolai} (for a review, see,
e.g., \cite{deWit} and references therein) and claims to the role
of M-theory Hamiltonian \cite{BFSS}. 
This conjecture and the fact
that the low-energy limit of the M-theory is described by
eleven-dimensional supergravity pose the important question of
existence of zero-energy normalizable eigenfunctions. 
Using a complete set of gauge invariant variables and generalization of
the Born-Oppenheimer approximation the simplest case of $SU(2)$
matrix theory has been investigated and an asymptotic form of the
ground state was proposed \cite{HalpernSchwartz}
\footnote{
The case of the $SU(n)$ group with arbitrary $n \geq 2$ was considered in
\cite{Konechny,HaslerHoppe}.}.
Even the simplest of these dimensionally reduced models are still
rather complicated and possesses non-trivial dynamics. 
It was found
\cite{MatinyanSavvidyTer-Arutyunyan-Savvidy,AsatryanSavvidy,Medvedev}
that the classical non-supersymmetric $SU(2)$ Yang-Mills mechanics
exhibits chaotic behavior when the dynamics takes place on a
special invariant submanifold. 
It was proved that on this submanifold there is no analytical integral 
of motion except the
energy integral, and thus the Yang-Mills mechanics represents a
non-integrable system \cite{YMMechanics:Chaos}. 
A similar investigation of the classical dynamics of bosonic membrane matrix
model yielded again chaotic behavior. 
However, recently, in \cite{Arefeva,Arefeva-Spectrum} the supersymmetric 
$SU(2) \times SO(2)$ matrix model was investigated in detail and it was
demonstrated that there exists a chaos-order transition depending
on the value of the angular momentum.

In the present paper we shall continue the study of models
obtained from the $SU(2)$ Yang-Mills field theory under the
supposition of fields homogeneity. 
We consider the model of light-cone $SU(2)$ Yang-Mills 
classical mechanics and address the
problem of its complete Hamiltonian reduction and integrability.
Analogously to the instant form of Yang-Mills mechanics, the
light-cone version follows from the light-cone Yang-Mills field
theory when the gauge fields depend on the light-cone time only.
Both dynamical systems, obtained under such suppositions, contain
a finite number of degrees of freedom and inherit in a specific
form the gauge invariance of the original Yang-Mills theory. 
In a recent article we outlined such a difference  of the light-cone
version of Yang-Mills mechanics to its instant form counterpart
even in the character of the local gauge invariance
\cite{GKM-CAAP}. 
Now we present a result of the Hamiltonian
reduction of the light-cone $SU(2)$ Yang-Mills mechanics and
demonstrate that after elimination of all ignorable coordinates
the corresponding unconstrained Hamiltonian system represents a
simple integrable system.

We start with the formulation of the $SU(2)$ light-cone mechanics
as a degenerate Lagrangian model for a matrix valued variable $A$,
perform the standard Hamiltonian analysis proving that the
presence of constraints force the classical dynamics to develop on
the subspace of matrices with $\mathrm{rank}||A||=1\,$. 
Using the
adapted coordinates frame we show that it is equivalent to the
dynamics of a {\em free} particle in one dimension. We also study
the {\em complex solutions} to the second class constraints and
demonstrate that in this case the reduced system coincides with
the well-known model of so-called conformal mechanics, introduced
by V. de Alfaro, S. Fubini and G. Furlan
\cite{deAlfaroFubiniFurlan}.

After Dirac's famous paper \cite{Dirac1949} on different forms of
relativistic dynamics it has been recognized that the different
choice of the time evolution parameter can drastically change the
content and interpretation of the theory. The present study shows
that the long-wavelength approximation in instant and light-front
formulation leads to the models that differ drastically even in
sense of their classical integrability. 
The question whether models with the higher order gauge groups as well as after
inclusion of an additional supersymmetry stay integrable is still
open. 
It is also interesting to study the question of their
correspondence to the known superconformal generalizations of
conformal mechanics.
\footnote{ 
The $N=2$ supersymmetric extension of conformal
mechanics was generalized in \cite{AkulovPashnev,FubiniRabinovici}
to an $SU(1,1\mid 1)$ invariant superconformal mechanics. 
Soon after, $N=4$ extension of conformal mechanics with $SU(1,1\mid 1)$
superconformal symmetry was elaborated
\cite{Ivanov:N4,Ivanov:N4-N2} and using the geometric method the
superconformal mechanics was formulated in a manifestly invariant
manner for an arbitrary even $N$ \cite{Ivanov:N4}.}
Here we note, that the quantum mechanical model with periodicity
in  light-cone time, obtained by the dimensional reduction of the
light-cone version of $N=1$ super $SU(n)$ Yang-Mills theory was
studied in \cite{Kostov}.
It was shown that this model is
integrable in the sense that its partition function is a
tau-function of the Toda hierarchy and only in the large $n$ limit
can be solved exactly.

The fact that a finite dimensional system obtained by dimensional
reduction inherits the conformal symmetry of the original field
theory is not quite unexpected. 
One-dimensional conformally
invariant systems already appeared in black hole physics
\cite{BH-SCM} and cosmology \cite{PiolineWaldron}. 
However, to our knowledge, their relation to the light-cone Yang-Mills theory has
not been pointed out yet.

The organization of the rest of the paper is as follows. In
Section 2 we start with the Lagrangian formulation of the
light-cone model and give the standard analysis of Hamiltonian
constraints including their separation into the first and second
class constraints sets. 
Then in Section 3 the Hamiltonian reduction is performed. 
First the constraints generating the
$SU(2)$ gauge transformations are eliminated using the coordinates
adapted to the gauge symmetry. 
Further to this the reduction due
to the remaining first and second class constraints is carried out
exploiting the new convenient set of coordinates. Section 4 gives
our final conclusions and comments. 
The appendix is devoted to the
derivation of the Lagrangian equations of motion of the
unconstrained system starting from the Lagrangian equations of
motion for the light-cone $SU(2)$ Yang-Mills mechanics by
elimination of all Lagrangian constraints.


\section{Light-cone model and analysis of constraints}


In this Section we give the formulation of the $SU(2)$ light-cone
Yang-Mills mechanics, calculate all constraints and separate them
into the first and second class ones.


\subsection{Model formulation}


We start with the action of Yang-Mills field theory in
four-dimensional Minkowski space $M_4$, endowed with a metric
$\eta$ and represented in the coordinate free form
\begin{equation}
\label{eq:gaction}
I : = \frac{1}{g^2}\, \int_{M_4} \mbox{tr} \,  F\wedge * F \,,
\end{equation}
where $g$ is a coupling constant and the $su(2)$ algebra valued curvature two-form
\begin{equation}
F:= d A +  A \wedge A
\end{equation}
is constructed from the connection one-form $A$.
The connection and curvature, as Lie algebra valued quantities, are expressed in terms
of the antihermitian $su(2)$ algebra basis $\tau^a = \sigma^a/2 i$
with the Pauli matrices $\sigma^a \,, a = 1,2,3$,
\begin{equation}
A = A^a \, \tau^a \,, \qquad F = F^a \, \tau^a\,\,.
\end{equation}
The metric $\eta$ enters the action through the dual field
strength tensor defined in accordance with the Hodge star
operation
\begin{equation}
* F_{\mu\nu}  = \frac{1}{2}\,\sqrt\eta\, \epsilon_{\mu\nu\alpha\beta}\,
F^{\alpha\beta}\,.
\end{equation}

To formulate the light-cone version of the theory let us introduce
the basis vectors in the tangent space $T_P(M_4)$
\begin{equation}
e_{\pm} := \frac{1}{\sqrt 2} \, \left( e_0 \pm e_3 \right) \,, \quad
e_\bot :=  \left( e_k \,, k = 1, 2 \right) \,.
\end{equation}
The first two vectors are tangent to the light-cone and the
corresponding coordinates are referred usually as the light-cone
coordinates $ x^\mu = \left( x^+, x^-, x^\bot\right) $
\begin{equation}
x^\pm := \frac{1}{\sqrt 2}\, \left( x^0 \pm x^3 \right) \,, \qquad
x^\bot :=   x^k \,,\quad k = 1, 2 \,.
\end{equation}
The non-zero components of the metric $\eta$ in the light-cone
basis $\left(e_+, e_-, e_k \right)$ are
\begin{equation}
\eta_{+-} = \eta_{-+} = - \eta_{11} = - \eta_{22} = 1\,.
\end{equation}
The connection one-form in the light-cone basis is given as
\begin{equation} \label{eq:conlc}
A = A_+ \, dx^+ + A_- \, dx^- + A_k \, dx^k \,.
\end{equation}

By definition the Lagrangian of light-cone Yang-Mills mechanics
follows from the corresponding Lagrangian of Yang-Mills theory if
one supposes that the components of the connection one-form $A$ in
(\ref{eq:conlc}) depend on the light-cone ``time variable'' $x^+$
alone
\begin{equation}
A_\pm = A_\pm(x^+)\,, \qquad A_k = A_k(x^+) \,.
\end{equation}
Substitution this {\em ansatz} into the classical action
(\ref{eq:gaction}) defines the Lagrangian of light-cone Yang-Mills
mechanics
\begin{equation} \label{eq:lagr}
L =
\frac{1}{2g^2} \, \left( F^a_{+ -} \,  F^a_{+ -} +
2 \, F^a_{+ k} \, F^a_{- k} - F^a_{12} \, F^a_{12} \right)\,,
\end{equation}
where the light-cone components of the field-strength tensor are given
by
\begin{eqnarray}
&& F^a_{+ -} = \frac{\partial A^a_-}{\partial x^+} + \epsilon^{abc}\, A^b_+ \,  A^c_- \,,\\
&& F^a_{+ k} = \frac{\partial A^a_k}{\partial x^+} + \epsilon^{abc}\, A^b_+ \,  A^c_k \,, \\
&& F^a_{- k} = \epsilon^{abc} \,  A^b_- \, A^c_k \,, \\
&& F^a_{i j} = \epsilon^{abc}\, A^b_i \, A^c_j\,, \quad i,j,k = 1,2 \,.
\end{eqnarray}
Hence, the Yang-Mills light-cone mechanics is a finite
dimensional system with configuration coordinates
$A_\pm \,,A_k$ whose evolution with respect to the time $\tau$
\begin{equation}
\tau := x^+\,
\end{equation}
is determined by the Lagrangian (\ref{eq:lagr}).


\subsection{Generalized Hamiltonian dynamics}


Performing the Legendre transformation
\footnote{
To simplify the formulas we shall use overdot to denote
derivative of a function with respect to light-cone time  $\tau$.
Further, we shall treat in equal footing the up and down
isotopic indexes denoted with $a, b, c, d$.}
\begin{eqnarray}
&&
\pi^+_a  =  \frac{\partial L}{\partial \dot{A^a_+}} =0\,,\\
&&
\pi^-_a  =  \frac{\partial L}{\partial \dot{A^a_-}} =
\frac{1}{g^2} \, \left( \dot{A^a_- } + \epsilon^{abc} \, A^b_+ \, A^c_- \right) \,, \\
&&
\pi_a^k = \frac{\partial L}{\partial \dot{A^a_k}} =
\frac{1}{g^2} \, \epsilon^{abc} \, A^b_- \, A^c_k \,,
\end{eqnarray}
we obtain the canonical Hamiltonian
\begin{equation} \label{eq:khrlh}
H_C = \frac{g^2}{2}\,  \pi^-_a  \pi^-_a - \,
\epsilon^{abc} \,  A^b_+ \left(A^c_- \, \pi^-_a  + A^c_k \,\pi^k_a \right) + V(A_k)
\end{equation}
with a potential term
\begin{equation}
V(A_k) = \frac{1}{2 g^2} \,
\left[
\left(A^b_1 A^b_1\right)\, \left(A^c_2 A^c_2 \right) -
\left(A^b_1 A^b_2\right)\, \left(A^c_1 A^c_2 \right)
\right] \,.
\end{equation}
The non-vanishing Poisson brackets between the fundamental canonical variables are
\begin{eqnarray}
&& \{ A^a_\pm \,, \pi^\pm_b \} = \delta^a_b \,,\\
&& \{ A^a_k \,,  \pi_b^l \} = \delta_k^l \delta^a_b \,.
\end{eqnarray}

The Hessian of the Lagrangian system (\ref{eq:lagr}) is
degenerate, $\det ||\frac{\partial^2 L}{\partial \dot{A}\partial
\dot{A}}||= 0$, and as a result there are
primary constraints
\begin{eqnarray}
&& \varphi^{(1)}_a := \pi^+_a = 0 \,, \label{eq:prcon-1} \\
&& \chi^a_k := g^2 \, \pi^a_k  + \epsilon^{abc} \, A^b_- A^c_k =
0\,, \label{eq:prcon2}
\end{eqnarray}
satisfying the following Poisson brackets relations
\begin{eqnarray}
&& \{ \varphi^{(1)}_a \,, \varphi^{(1)}_b\} = 0 \,,\\
&& \{ \varphi^{(1)}_a \,, \chi^b_k \} = 0 \,,\\
&& \{ \chi^a_i \,, \chi^b_j\} = -2\, g^2 \epsilon^{abc}\, A^c_- \,
\eta_{i j} \,. \label{eq:prcon-2}
\end{eqnarray}

According to the Dirac prescription, the presence of primary
constraints affects the  dynamics of the degenerate system. Now
the generic evolution is governed by the total Hamiltonian
\begin{equation} \label{eq:toth}
H_T = H_C + U_a(\tau)\varphi^{(1)}_a + V^a_k(\tau)\chi^a_k\,,
\end{equation}
where $U_a(\tau)$ and $V^a_k(\tau)$ are unspecified functions of
the light-cone time $\tau$.
Using this Hamiltonian the dynamical
self-consistence of the primary constraints may be checked. From
the requirement of conservation of the primary constraints
$\varphi^{(1)}_a$ it follows
\begin{equation}
0 = \dot \varphi^{(1)}_a = \{\pi^+_a\,, H_T\} =
\epsilon^{abc} \left(A^b_-  \pi^-_c  +  A^b_k \pi^k_c \right)\,.
\end{equation}
Therefore there are three secondary constraints $\varphi^{(2)}_a$
\begin{equation} \label{eq:secgauss}
\varphi^{(2)}_a := \epsilon_{abc}
\left(A^b_-  \pi^-_c  +  A^b_k \pi^k_c \right)=0\,,
\end{equation}
which obey the $so(3, \mathbb{R})$ algebra
\begin{equation}
\{ \varphi^{(2)}_a \,, \varphi^{(2)}_b \} = \epsilon_{abc}\, \varphi^{(2)}_c \,.
\end{equation}
The same procedure for the primary constraints $\chi^a_k$ gives
the following self-consistency conditions
\begin{equation} \label{eq:secchi}
0 = {\dot\chi}^a_k =
\{\chi^a_k\,, H_C \} - 2\, g^2\, \epsilon^{abc} \, V^b_k\, A^c_-   \,.
\end{equation}
The analysis of these equations depends on the properties of the
matrix $\mathcal{C}_{ab} = \epsilon^{abc}\, A^c_-$. This matrix is
degenerate with a rank varying from $0$ to $2$ depending on the
point of the configuration space. 
If its rank is $2$ then among
the six primary constraints $\chi^a_k$ there are two first class
constraints and a maximum of four Lagrange multipliers $V^b_k$ can
be determined from (\ref{eq:secchi}). 
When the rank of the matrix
$\mathcal{C}_{ab}$ is minimal, the locus points are $A^a_-=0$ and
all six constraints $\chi^a_k$ are Abelian ones. 
For such an exceptional configuration the constrained system reduces to the
dynamically trivial one.  
Hereinafter we shall consider the
subspace of configuration space where ${\rm rank}||\mathcal{C}||=2$. 
For those configurations we are able to
introduce the unit vector
\begin{equation}
N^a = \frac{A^a_-}{ \sqrt{(A_-^1)^2 + (A_-^2)^2 + (A_-^3)^2} }\,,
\end{equation}
which  is a null vector of the matrix $ \| \, \epsilon^{abc} \,
A^c_- \,\|,$ and to decompose the set of six primary constraints
$\chi^a_k$ as
\begin{eqnarray}\label{eq:abpsi}
&&
\psi_k : = N^a \chi^a_k  \,,\\
&&
\chi^a_{k\bot} :=
\chi^a_k - \left( N^b \chi^b_k  \right) \, N^a \,.
\end{eqnarray}
In this decomposition the first two constraints $\psi_k$ are
functionally independent and satisfy the Abelian algebra
\begin{equation}
\{ \psi_i \,, \psi_j \} = 0 \,,
\end{equation}
while the constraints $\chi^a_{k \bot}$ are functionally dependent
due to the conditions
\begin{equation} \label{eq:dependence}
N^a \, \chi^a_{k \bot} = 0\,.
\end{equation}
Choosing among them any four independent constraints we can
determine four Lagrange multipliers $V^k_{\ b \bot}$.

The Poisson brackets of the constraints $\psi_k$ and $\varphi^{(2)}_a$
with the total Hamiltonian vanish after projection on the
constraint surface (CS) defined by equations $\psi_k=0$ and
$\varphi^{(2)}_a=0$
\begin{eqnarray} \label{eq:check}
&&
\{ \psi_k \,, H_T \}{\,\vert_{CS}} = 0 \,, \\
&&
\{ \varphi^{(2)}_a \,, H_T \}{\,\vert_{CS}} = 0\,
\end{eqnarray}
and thus there are no ternary constraints.


Summarizing, we arrive at the set of constraints $\varphi^{(1)}_a,
\psi_k, \varphi^{(2)}_a, \chi^b_{k \bot}$. The Poisson brackets
algebra of the first three is
\begin{eqnarray}
&&
\{ \varphi^{(1)}_a \,, \varphi^{(1)}_a\} = 0 \,,
\label{eq:group_1} \\
&&
\{ \psi_i \,, \psi_j \} = 0 \,,
\label{eq:group_2} \\
&&
\{ \varphi^{(2)}_a \,, \varphi^{(2)}_b\} = \epsilon_{abc}\, \varphi^{(2)}_c \,,
\label{eq:group_3} \\
&&
\{ \varphi^{(1)}_a \,, \psi_k\} =  \{\varphi^{(1)}_a \,,\varphi^{(2)}_b \} =
\{ \psi_k \,, \varphi^{(2)}_a \} = 0 \,.
\label{eq:group_4}
\end{eqnarray}
The constraints $\chi^b_{k \bot}$ satisfy the relations
\begin{equation} \label{eq:bracket-1}
\{ \chi^a_{i \bot} \,, \chi^b_{j \bot} \} = - 2 \, g^2 \, \epsilon^{abc}
\, A^c_- \, \eta_{i j}\,,
\end{equation}
and the Poisson brackets between these two sets of constraints are
\begin{eqnarray}
&&
\{\varphi^{(2)}_a \,, \chi^b_{k \bot} \} =
\epsilon^{abc} \, \chi^c_{k \bot} \,, \label{eq:bracket-2}\\
&& \{ \varphi^{(1)}_a \,, \chi^b_{k \bot}\} =
\{ \psi_i \,, \chi^b_{j\bot}\} = 0 \,. \label{eq:bracket-3}
\end{eqnarray}

From these relations we conclude that the model has 8 first-class
constraints $\varphi^{(1)}_a, \psi_k, \varphi^{(2)}_a $ and 4
second-class constraints $\chi^a_{k \bot}$. Counting the degrees
of freedom taking into account all these constraints, we obtain
that instead of $24$ constrained phase space degrees of freedom
there are $24 - 2 (5 + 3) - 4 = 4$ unconstrained degrees of
freedom, in contrast to the instant form of Yang-Mills mechanics
where the number of the unconstrained canonical variables is $12$.


\section{Unconstrained version of light-cone mechanics}


Now we shall perform a Hamiltonian reduction of the degrees of
freedom starting with an elimination of the gauge degrees of
freedom associated to the $SU(2)$ constraints $\varphi^{(2)}_a$.
The purpose of the present part of the paper is to rewrite the
theory in terms of special coordinates adapted to the action of
this gauge symmetry.


\subsection{Polar decomposition}


Let us organize the configuration variables $A^a_i$ and $A^a_-$ in
one $3\times 3$ matrix $A_{ab}$ whose entries of the first two
columns are $A^a_i$ and third column is composed by the elements $A^a_-$
\begin{equation} \label{eq:amatr}
A_{ab} : = \| A^a_1\,, A^a_2\,, A^a_-\|\,,
\end{equation}
and the momentum variables similarly
\begin{equation}\label{eq:mommat}
\Pi_{ab}:= \| \pi^{a1}\,, \pi^{a2}\,, \pi^{a-}\|.
\end{equation}
In order to find an explicit parametrization of the orbits with
respect to the gauge symmetry action, it is convenient to use a
polar decomposition \cite{Zelobenko} for the matrix $A_{ab}$
\begin{equation}\label{eq:polar}
A = O S\,,
\end{equation}
where $S$ is a positive definite $3 \times 3$ symmetric matrix,
$O(\phi_1,\phi_2, \phi_3) = e^{\phi_1 J_3}e^{\phi_2 J_1}e^{\phi_3
J_3}$ is an orthogonal matrix parameterized by the three Euler
angles $(\phi_1,\phi_2,\phi_3)$. 
The matrices  $(J_a)_{ij} = \epsilon_{iaj}$ are the 
$SO(3, \mathbb{R})$ generators in adjoint
representation.

It is in order to make a few remarks on the change of variables in
(\ref{eq:polar}). 
It is well-known that the polar decomposition is
valid for an arbitrary matrix. 
However, the orthogonal matrix in
(\ref{eq:polar}) is uniquely determined only for an invertible
matrix $A$
\begin{equation}
\label{eq:f0s} O = A S^{-1 }\,, \quad\quad S=\sqrt{A A^T}~.
\end{equation}
The non-degenerate $3\times 3$ matrices can be identified with an
open set of the $\mathbb{R}^9$ using the entries of the matrix
$A_{ab}$ as corresponding Cartesian coordinates and in this case
the polar decomposition (\ref{eq:polar}) is a uniquely invertible
transformation from these Cartesian coordinates to a new set of
coordinates, the entries of positive matrix $S$ and the angles
parameterized the orthogonal matrix $O$. 
For degenerate matrices a
more sophisticated analysis is necessary. 
Here we note only that the set of $n \times n$ matrices 
with rank $k$ is a manifold with
dimension $k(2n-k)$, but in contrast the no-degenerate case the
manifold atlas now necessarily contains several charts. 
Hence, for degenerate matrices $A$ the representation (\ref{eq:polar}) has to
be replaced by a more elaborated construction.

Now we shall limit ourselves to the subspace of non-degenerate
matrices and hence one can treat the polar decomposition
(\ref{eq:polar}) as a uniquely invertible transformation from the
configuration variables $A_{ab}$ to a new set of Lagrangian
variables: six coordinates $S_{ij}$ and three coordinates
$\phi_i$. It is worth to note here that in virtue of the
constraints (\ref{eq:prcon2}) the determinant of the matrix $A$ is
related to the third component of the gauge field spin
\begin{equation}\label{eq:detc}
2 \det A - g^2 \epsilon_{3ik}\,A^a_k\, \pi^a_i=0\,.
\end{equation}

The polar decomposition (\ref{eq:polar}) induces the point
canonical transformation from the coordinates $A_{ab}$ and
$\Pi_{ab}$ to new canonical pairs
$(S_{ab}, P_{ab})$ and $(\phi_a, P_a)$
with the following non-vanishing Poisson brackets
\begin{eqnarray}
&&\{ S_{ab} \,, P_{cd} \} =
\frac{1}{2}
\left( \delta_{ac}\, \delta_{bd} + \delta_{ad} \, \delta_{bc} \right)\,,\\
&&\{ \phi_{a} \,, P_{b} \} =  \delta_{ab}\,.
\end{eqnarray}
The expression of the old $\Pi_{ab}$ as a function of the new
coordinates is
\cite{AKDM:PLA2002,AKDM:PAN2002}
\begin{equation} \label{eq:polmom}
\Pi = O \left( P - k_a J_a \right)\,,
\end{equation}
where
\begin{equation}
k_a = \gamma^{-1}_{ab} \left(\eta^L_b - \varepsilon_{bmn}
\left(S P \right)_{mn}  \right)\,,
\end{equation}
$ \gamma_{ik} = S_{ik} - \delta_{ik}\, \mbox{tr} S $ and
$\eta^L_a$ are three left-invariant vector fields on the $\sot$
group
\begin{eqnarray}
&&\label{eq:liv1}
\eta^L_1 =
\frac{ \sin\phi_3 }{\sin\phi_2 }\,  P_1 +
\cos\phi_3 \,  P_2 -
\cot\phi_2 \sin\phi_3 \  P_3 \,, \\
&&\label{eq:liv2}
\eta^L_2 =
\frac{ \cos\phi_3 }{ \sin\phi_2 }\,  P_1 -
\sin\phi_3 \,  P_2 - \cot\phi_2 \cos\phi_3 \ P_3 \,,\\
&&\label{eq:liv3}
\eta^L_3 = P_3\,.
\end{eqnarray}

In terms of the new variables the constraints take the form
\begin{eqnarray}
&& \label{eq:gconssGL1}
\varphi^{(2)}_a = O_{ab}\, \eta^L_b\,,\\
&&\label{eq:conssSC2}
\chi_{am} =
O_{ab} \, \left(P_{bm} + \epsilon_{bmc}\,k_c + \epsilon_{bij}\, S_{i3}\, S_{jm}\right)\,.
\end{eqnarray}
Thus one can pass to the equivalent set of constraints
\begin{eqnarray}\label{eq:gc}
&&\eta^L_a = 0\,, \\
&&\widetilde{\chi}_{ai} = P_{ai} + \epsilon_{aij}\,
\gamma^{-1}_{jk}\ \epsilon_{kmn}(SP)_{mn}
 + \epsilon_{amn}\, S_{m3}\, S_{ni}=  0\,\label{eq:gsc}
\end{eqnarray}
with vanishing Poisson brackets
\begin{equation}\label{eq:nalg}
\{\eta^L_a, \widetilde{\chi}_{bi}\} = 0\,.
\end{equation}

Using the polar decomposition (\ref{eq:polar}) and (\ref{eq:polmom})
we separate the variables $(S_{ab},P_{ab})$, invariant under
gauge transformations generated by Gauss law constraints
$\varphi^{(2)}_a$, from the gauge variant ones $(\phi_a, P_a)$.
Now in order to eliminate all gauge degrees of freedom related to
this symmetry it is enough to project to the constraint shell
described by condition of nullity of the Killing vector fields $\eta_a^L$.
After projection the corresponding cyclic degrees of
freedom, the angles $\phi_a$, automatically disappear from the
projected Hamiltonian.


\subsection{Main-axes decomposition}


In order to proceed further in resolution of the remaining
constraints (\ref{eq:gsc}) we introduce the main-axes
decomposition for the symmetric $3\times 3$ matrix $S$
\begin{equation}\label{eq:mainax}
S =
R^T(\chi_1, \chi_2, \chi_3)
\left(
\begin{array}{ccc}
q_1   &   0    &    0 \\
0     &   q_2  &    0 \\
0     &   0    &    q_3
\end{array}
\right)
R(\chi_1, \chi_2, \chi_3)\,,
\end{equation}
with orthogonal matrix $R(\chi_1,\chi_2, \chi_3) = e^{\chi_1
J_3}e^{\chi_2 J_1}e^{\chi_3 J_3}$, parameterized by three Euler
angles $(\chi_1,\chi_2,\chi_3)$. The Jacobian of this
transformation is
\begin{equation}
\frac{\partial(\ S_{i<j}\ )}{\partial(q_a, \chi_b)} \sim \prod_{a
\neq b}^{3} \mid q_a - q_b \mid.
\end{equation}
Therefore equation (\ref{eq:mainax}) can be used as definition of
new configuration variables: three ``diagonal'' variables $(q_1,
q_2, q_3)$, eigenvalues of the matrix $S$, and three angular
variables $(\chi_1, \chi_2,\chi_3)$, if and only if all
eigenvalues of the matrix $S$ are different, 
$q_1 \neq q_2 \neq q_3\,$. 
The eigenvalues  $q_a$ parameterize the orbits of the
adjoint action of $\sot$ group in the space of $3 \times 3$
symmetric matrices and the configurations with $q_1 < q_2 < q_3$
represent the so-called principle orbit. Our consideration given
below is correct for this type of orbits whereas the treatment of
orbits with coinciding eigenvalues of the matrix $S$, the singular
orbits \cite{ORaifeartaigh}, requires different and more
elaborated treatment that is beyond the scope of the present
paper.

The momenta $p_a$ and $p_{\chi_a}$, canonically conjugated to the
diagonal $q_a$ and angular variables $\chi_a$, can be found using
the canonical invariance of the symplectic one-form
\begin{equation}
\sum^3_{a, b = 1}\, P_{ab} \, d S_{ab} =
\sum^3_{a = 1} \, p_a\, d q_a  + \sum^3_{a = 1}\, p_{\chi_a}\, d {\chi}_a \,.
\end{equation}
The original momenta $P_{ab}$, expressed in terms of the new canonical variables,
read
\begin{eqnarray} \label{eq:newmom}
P =
R^T \sum_{a = 1}^3
\left(
p_a \, \overline{\alpha}_a + {\cal P}_a \,\alpha_a
\right) R\,.
\end{eqnarray}
Here $\overline{\alpha}_a$ and $\alpha_a$ denote the diagonal and 
off-diagonal basis elements
of the space of symmetric matrices with orthogonality relations
\begin{equation}
\mbox{tr}\ (\overline{\alpha}_a \overline{\alpha}_b) = \delta_{ab}\,, \qquad
\mbox{tr}\ ({\alpha}_a {\alpha}_b) = 2 \delta_{ab}\,, \qquad
\mbox{tr}\ (\overline{\alpha}_a {\alpha}_b) = 0
\end{equation}
and
\begin{equation}
{\cal P}_a = - \frac{1}{2}\, \frac{\xi^R_a}{q_b - q_c}\,
\,\,\,\,\, (\mbox{cyclic permutations} \, a\not = b \not = c)\,.
\end{equation}

The $\xi^R_a $ are three $\sot$ right-invariant vector fields
given in terms of the angles $\chi_a$ and their conjugated momenta
$p_{\chi_a}$ via
\begin{equation}
\xi^R_a = M^{-1}_{ba} p_{\chi_b}\,,
\end{equation}
where the matrix $M$ is given by
\begin{equation} \label{eq:MCr}
M_{ab} = - \frac{1}{2}\,\mbox{tr} \left(J_a \frac{\partial R}{\partial\chi_b}
\, R^T\right)\,.
\end{equation}
The explicit form of the three $\sot$
right-invariant Killing vector fields is
\begin{eqnarray}
&& \label{eq:rf1}
\xi^R_1 =
 - \sin\chi_1 \cot\chi_2 \ p_{\chi_1} +
\cos\chi_1 \  p_{\chi_2} +
\frac{\sin\chi_1}{\sin\chi_2}\ p_{\chi_3} \,,\\
&& \label{eq:rf2}
\xi^R_2 =
\,\,\cos\chi_1 \cot\chi_2 \ p_{\chi_1} + \sin\chi_1 \  p_{\chi_2} -
\frac{\cos\chi_1}{\sin\chi_2}\ p_{\chi_3} \,, \\
&& \label{eq:rf3}
\xi^R_3 =  p_{\chi_1}\,.
\end{eqnarray}
Using these formulas the constraints $\widetilde{\chi}$ in
(\ref{eq:gsc}) may be rewritten in terms of the main-axes
variables as
\begin{eqnarray} \label{eq:mcons}
\widetilde{\chi} = \sum_{a = 1}^3\, R^T
\left[
\pi_a \, \overline{\alpha}_a - \frac{1}{2}\, \rho^-_a \alpha_a +
\frac{1}{2}\, \rho^+_a\, J_a \right] R\,,
\end{eqnarray}
where
\begin{equation}
\rho^\pm _a = \frac{\xi^R_a}{q_b\ \pm q_c} \pm \frac{1}{g^2}\, q_a n_a(q_b\ \pm \ q_c)\,,
\end{equation}
and $n_a = R_{a3}$.

Note that the constraint (\ref{eq:detc}) on the determinant of the
matrix $A$ now takes the form
\begin{equation}\label{eq:detconstr}
2\,q_1\,q_2\,q_3 - g^2 \xi_3^L = 0\,,
\end{equation}
where $\xi_3^L$ is the third left-invariant Killing vector field,
$\xi^L_a = R_{ab}\, \xi^R_{b}$
\begin{eqnarray} &&\label{eq:lixi}
\xi^L_1 = \frac{ \sin\chi_3 }{\sin\chi_2 }\,  p_{\chi_1} +
\cos\chi_3 \, p_{\chi_2} -
\cot\chi_2 \sin\chi_3 \  p_{\chi_3} \,, \\
&&\xi^L_2 = \frac{ \cos\chi_3 }{ \sin\chi_2 }\, p_{\chi_1} -
\sin\chi_3 \,  p_{\chi_2} - \cot\chi_2 \cos\chi_3 \ p_{\chi_3} \,,\\
&&\xi^L_3 = p_{\chi_3}\,.
\end{eqnarray}

As was shown above the constraints $\chi^a_i$ represent the mixed
system of first and second class constraints $\psi_i$ and
${\chi^a_i}_\bot$. 
To perform the reduction to the constraint
shell it is useful at first to introduce the gauge fixing
condition and eliminate the two first class constraints $\psi_i$.
The expression (\ref{eq:abpsi}) for the Abelian constraints
$\psi_i$ dictates the appropriate gauge fixing condition
\begin{equation}\label{eq:gcon}
\overline{\psi}_i := N^a A^a_i = 0\,,
\end{equation}
which is the canonical one in the sense that
\begin{equation}\label{eq:cgc}
\{\overline{\psi}_i,\ \psi_j\} \ =\  \delta_{ij}\,.
\end{equation}
The constraints  $\psi_i=0 $ rewritten in terms of the main-axes
variables may be identified with the nullity of the momenta
\begin{equation}\label{eq:psima}
p_{\chi_1} = 0\,, \qquad p_{\chi_2} = 0\,,
\end{equation}
while  the canonical gauge-fixing condition (\ref{eq:gcon}) fixes
the corresponding angular variables $\chi_1$ and $\chi_2$
\begin{equation}\label{eq:magf}
\chi_1 = \frac{\pi}{2}\,, \qquad \chi_2 = \frac{\pi}{2}\,.
\end{equation}

Introduction of the gauge fixing conditions (\ref{eq:magf}) means
that all constraints are now second class ones and therefore the
reduction to unconstrained variables can now  be achieved by the
projection of canonical Hamiltonian onto the constraint shell with
simultaneously replacement of the canonical Poisson brackets by
the Dirac ones.

Projection of the canonical Hamiltonian (\ref{eq:khrlh}) to the
surface described by constraints (\ref{eq:psima}) and
(\ref{eq:magf}) gives
\begin{equation}\label{eq:hymcm}
H_{LC}:=H_{C}( \chi_1=\frac{\pi}{2}\,, p_{\chi_1} =0\,,
\chi_2=\frac{\pi}{2}\,, p_{\chi_2} =0 ) = \
\frac{g^2}{2}\left(p^2_1\ + \frac{q_2^2\, q_3^2}{g^4}\right)\,.
\end{equation}
Furthermore, taking into account the constraint
(\ref{eq:detconstr}) the projected Hamiltonian (\ref{eq:hymcm})
may be rewritten as
\begin{equation}\label{eq:hymcmp}
H_{LC}\biggl |_{2\,q_1\,q_2\,q_3 - g^2 \xi_3^L = 0\,}\ = \
\frac{g^2}{2}\left( p^2_1\ + \left(\frac{\xi^L_3}{2q_1}\right)^2
\right)\,.
\end{equation}

But it is not the end of the reduction procedure. 
Two further steps are required. 
First, it is necessary to examine all four
second class constraints ${\chi^a_i}_\bot$ and to verify whether
(\ref{eq:hymcmp}) is indeed the expression for the reduced
Hamiltonian describing the dynamics of unconstrained variables.
Second, it is necessary to calculate the fundamental Dirac
brackets between unconstrained variables in order to determine the
correct equation of motion.

It may be checked that the constraints ${\chi^a_i}_\bot$ lead to
the conditions on the ``diagonal'' canonical pairs $(q_i\,,p_i)$.
Namely, the canonical momenta $p_2$ and $p_3$ are vanishing
\begin{eqnarray}\label{eq:sma1}
p_2 =0 \,, \quad p_3=0\,,
\end{eqnarray}
while the corresponding coordinates $q_2$ and $q_3$ are subject to
the constraint
\begin{equation}\label{eq:fc}
q_2^2 + q_3^2 = 0
\end{equation}
as well the constraint (\ref{eq:detconstr}). The {\em real}
solution of the equation (\ref{eq:fc}) is the trivial one $q_2=
q_3 = 0$. For this solution according to the constraint
(\ref{eq:detconstr}) $\xi^L_3$ turns to be zero and thus the
Hamiltonian (\ref{eq:hymcmp}) reduces further to a Hamiltonian of
free one-dimensional particle motion.

Here it is in order to make an explanatory comment, because we
arrived at certain contradiction to our initial assumptions. The
reduced Hamiltonian system obtained here contains only two degrees
of freedom, while according to the counting given at the end of
the Section 2, we were expected to obtain a 4-dimensional
unconstrained system. 
In order to explain this contradiction note
that this counting was based on the assumption that the
configuration space of the initial Lagrangian system is
$12$-dimensional or in another words the $3\times 3$ matrix $A$ in
(\ref{eq:amatr}) is non-degenerate. 
However the constraint
(\ref{eq:fc}) states that $\mathrm{det}|| A ||=\mathrm{det}|| S
||=0\,,$ and rigorously speaking our consideration shows only that
the dynamics of the unconstrained system develops on the subspace
with $\mathrm{rank}|| A ||\leq 2$. 
Therefore it is necessary to
consider the configuration space of the initial Lagrangian system
consisting from the degenerate matrices with rank less than
maximal and perform the whole analysis again.
\footnote {
For example, the counting of the degrees of freedom is now as follows:
the dimension of subspace of
$3\times3 $ matrices with rank $k=2$ is $2\times(2\times3-2)=8$.
So, the configuration space of the initial system is not $12$-dimensional, but
$11$-dimensional and thus the reduced Hamiltonian system contains
only $22 - 2 (5 + 3) - 4 = 2$ degrees of freedom.}

However instead of an explicit parametrization of the
configuration space with $\mathrm{rank}|| A || = 2$ and
$\mathrm{rank}|| A || = 1$ we use the following trick.
\footnote{
As justification, in the appendix we give an alternative
derivation directly from the Lagrangian equation of motion by
solving the corresponding Lagrangian constraints.}
Let us consider
the analytic continuation of the constraint (\ref{eq:fc}) into a
complex domain and explore its {\em complex } solution
\begin{equation}\label{eq:cs}
q_2= \pm\, i\, q_3\,.
\end{equation}
Expressing $q_3$ from equation (\ref{eq:detconstr})
\begin{equation}\label{eq:imq}
q_3 = \frac{1 \mp i}{2}\, \sqrt{\frac{g^2 \xi^L_3}{q_1}}\,,
\end{equation}
we find that  $(q_1,p_1)$ and $(\chi_3, p_{\chi_3})$ remain {\em
real} unconstrained variables whose Dirac brackets are the
canonical ones
\begin{equation}\label{eq:DB}
\{q_1, p_1\}_D =1\,, \qquad \{\chi_3, p_{\chi_2}\}_D =1\,.
\end{equation}
Therefore the  dynamics of the unconstrained pairs $(q_1,p_1)$ and
$(\chi_3, p_{\chi_3})$ is given by the standard Hamilton equations
with the Hamiltonian (\ref{eq:hymcmp}). 
Remarking that the $\xi^L_3$ is conserved we conclude that (\ref{eq:hymcmp})
coincides with the Hamiltonian of conformal mechanics
\begin{equation}\label{eq:rhym}
H = \ \frac{g^2}{2}\left( p^2_1\ +
\frac{\kappa^2}{q_1^2}\right)\,,
\end{equation}
with  ``coupling constant''
$\kappa^2 = \left({\xi^L_3}/{2}\right)^2\,$ determined by the value of the
gauge spin, while the gauge field coupling constant $g$ controls
the scale for the evolution parameter.

From equation (\ref{eq:imq}) it follows that the quantity $\kappa$
is the parameter which measures the deviation from the real
classical trajectories. 
They all are laying in subspace with
$\mathrm{det}||A||=0$ and are described as the integral curves of
the Hamiltonian (\ref{eq:hymcmp}) with vanishing coupling constant
$\kappa=0\,,$ and therefore indeed correspond to a free particle
motion.


\section{Concluding remarks}


To conclude, we have considered the light-cone $SU(2)$ Yang-Mills
field theory supposing that the gauge potentials in the classical
action are functions only of the light-cone time. 
As we have demonstrated this {\em ansatz} effectively reduces the field
theory to a degenerate Lagrangian mechanical system whose
unconstrained version significantly differs from the
corresponding well-known instant time Yang-Mills mechanics.
Comparing with the instant form dynamics, the light-cone version
of Yang-Mills mechanics has a more complicated description
considered as a constrained system. 
Applying the Dirac Hamiltonian
method, we found that now the constraint content of the theory is
richer: there is, apart from the expected constraints which are
generators of the $SU(2)$ gauge transformations, a new set of
first and second class constraints. 
The presence of the new
constraints leads to an essential decre of the number of the
``true'' degrees of freedom and finally to its integrability.

In the present paper we have studied the Hamiltonian reduction of
the degenerate light-cone Yang-Mills mechanical system but have
left open several related questions such as analysis of the
symmetries, both gauge and rigid ones. 
The knowledge of symmetries
allows to understand the roots of the classical integrability of
the system and we plan to give the detailed presentation of these
investigations elsewhere.

We end this section with a remark about the possible link between
the classical integrability of the model obtained in
long-wavelength approximation of light-cone theory and the
properties of the corresponding vacuum. 
The nonintegrability and
chaotic nature of the instant form Yang-Mills mechanics is usually
treated as the manifestation of existence of the non-trivial
structure of the physical vacuum of gauge theories (see e.g.
\cite{Matinyan:Review1985}). 
On the other hand it is well-known
that owing to purely kinematical reasons the physical light-cone
vacuum of the theory coincides with the free Fock vacuum
\cite{Weinberg}. 
Therefore it seems that the integrability of the
light-cone Yang-Mills mechanics opposite to its instant
counterpart model is in accordance with the different vacuum
structures in these two forms of dynamics.


\section*{Acknowledgments}


Helpful discussions during the work on the paper with
I.Ya.~Aref'eva, B.~Dimitrov, G.~Gabadadze, E.A.~Ivanov,
A.S.~Koshelev, A.N.~Kvinikhidze, M.D.~Mateev, P.B.~Medve\-dev,
A.I.~Pashnev, V.P.~Pavlov, V.A.~Rubakov and O. Schroeder are
acknowledged. A.K. would like to thank the Abdus Salam
International Centre for Theoretical Physics (ICTP), Trieste,
Italy for support and kind hospitality extended to him in 2002 at
ICTP High Energy Section, where part of this study has been done.

This work was supported in part by the RFBR Grant No. 04-01-00784.
Contribution of V.G. was also partially supported by the grant
2339.2003.2 from the Russian Ministry of Science and Education.
A.K. acknowledges INTAS for providing financial support, Grant No.
00-00561.

\appendix

\section[]{Appendix: The Euler-Lagrange Equations}


The first variation of the  Lagrangian (\ref{eq:lagr}) with
respect to the variables $A^a_+\,, A^a_k$ and $A^a_- $ gives the
constraints, equations containing only first order derivatives,
and  the proper equations of motion.
Among the constraints there
are the Gauss Law equations ({\it summation over $k=1,2$})
\begin{equation}\label{eq:E-Lgl}
\dot{\bf{A}}_-\times{\bf{A}}_- +  {\bf{A}}_-({\bf{A}}_k\cdot
{\bf{A}}_k)- {\bf{A}}_k({\bf{A}}_k\cdot\bf{A}_-)-
{\bf{A}}_+(\bf{A}_-\cdot{\bf{A}}_-)+ {\bf{A}}_-
({\bf{A}}_+\cdot{\bf{A}}_-)=0\,,
\end{equation}
as well as the additional constraints ({\it no summation over
$i,k=1,2$ and $i\neq k$})
\begin{eqnarray}\label{eq:E-Lcc}
2\dot{\bf{A}}_k\times{\bf{A}}_-&-&\dot{\bf{A}}_-\times{\bf{A}}_k
-{\bf{A}}_k({\bf{A}}_i\cdot{\bf{A}}_i)+ {\bf{A}}_i({\bf{A}}_i\cdot
{\bf{A}}_k)\,\nonumber\\
&+&
2{\bf{A}}_k({\bf{A}}_+\cdot{\bf{A}}_-)-{\bf{A}}_-({\bf{A}}_+\cdot{\bf{A}}_k)
- {\bf{A}}_+({\bf{A}}_-\cdot{\bf{A}}_k)=0\,.
\end{eqnarray}
The true equations of motion, containing second order derivatives
of the variables ${\bf{A}}_-$ are ({\it summation over $k=1,2$})
\begin{eqnarray}
\ddot{\bf{A}}_- &+& \dot{\bf{A}}_+\times {\bf{A}}_- +
{\bf{A}}_+\times {\bf{\dot{A}}}_- +
{\bf{A}}_+\times{\bf{\dot{A}}}_-
-{\bf{A}}_k\times{\bf{\dot{A}}}_k\nonumber \\
& +&{\bf{A}}_+\left(
({\bf{A}}_+\cdot{\bf{A}}_-)-({\bf{A}}_k\cdot{\bf{A}}_k)\right)
-{\bf{A}}_-({\bf{A}}_+\cdot{\bf{A}}_+)+{\bf{A}}_k({\bf{A}}_+\cdot
{\bf{A}}_k)=0\,.
\label{eq:E-Le}
\end{eqnarray}
Here we have introduced the vector notation for the isotopic
components of the vector potential ${\bf{A}}_\pm =(A^1_\pm\,,
A^2_\pm\,, A^3_\pm)$ and ${\bf{A}}_k =(A^1_k\,, A^2_k\,, A^3_k)$.
The standard definitions for {\em dot} and {\em cross} product of
three dimensional isotopic vectors are used as well.

The aim of this Appendix is to show how to pass from this system
of nonlinear Euler-Lagrange equations
(\ref{eq:E-Lgl})-(\ref{eq:E-Le}) to the one-dimensional equation
of motion of a free particle for {em real} solutions and to
conformal mechanics for the case of {\em complex} solution to the
Lagrangian constraints.

To be close to the Hamiltonian consideration given in the main text
let us introduce in the isotopic space a positively oriented
orthonormal frame
of unit vectors, $({\bf l}\,,{\bf m}\,,{\bf n}\,)$,
\begin{eqnarray}\label{eq:btr}
{\bf{l}\cdot\bf{l}}&=&1\,,\qquad {\bf{l}}={\bf{m}\times \bf{n}}\,,\\
{\bf{m}\cdot\bf{m}}&=&1\,,\qquad {\bf{m}}={\bf{n}\times \bf{l}}\,,\\
{\bf{n}\cdot\bf{n}}&=&1\,,\qquad {\bf{n}}={\bf{l}\times \bf{m}}\,,
\end{eqnarray}
and start with the following {\em ansatz } for the  gauge
potential
\begin{eqnarray}\label{eq:gpa1}
    {\bf A}_- & = & x\,{\bf{n}}\,, \\
    {\bf A}_k & = & S_{1k}\,{\bf{l}} + S_{2k}\,{\bf{m}} \,.
    \label{eq:gpa2}
\end{eqnarray}
Note that this {\em ansatz} corresponds to the polar decomposition
(\ref{eq:polar}) when the only nonvanishing element in the third
row and the third column of the matrix $S$ is $S_{33}=x$ and
moreover it is supposed in (\ref{eq:gpa2}) that
\begin{equation}\label{eq:symm}
S_{12}=S_{21}\,.
\end{equation}
By construction the {\em ansatz} is such that the potentials
(\ref{eq:gpa1}) and (\ref{eq:gpa2}) obey four equations
\begin{equation}
{\bf{l}}\cdot{\bf{A}}_- = 0\,, \qquad
{\bf{m}}\cdot{\bf{A}}_- = 0\,,\qquad
{\bf{n}}\cdot{\bf{A}}_k=0 \,.
\end{equation}
Note that these equations are equivalent to the two primary
Abelian constraints (\ref{eq:abpsi}) and to the two gauge-fixing
conditions (\ref{eq:gcon}) imposed on the gauge potential in the
main text.

Now we shall demonstrate that the system of equations
(\ref{eq:E-Lgl})-(\ref{eq:E-Le}) admits a separation into three
subsets. The first one establish the connection between ${\bf
A}_+$ component of the gauge potential and the frame $({\bf
l}\,,{\bf m}\,,{\bf n}\,)$, the second one consists of the
equations for the variables $S_{1k}$ and $S_{2k}$ and the third
one represents only one second order differential equation for the
variable $x$ with a parameter, whose value is the first integral
of the equations for the variables $S_{1k}$ and $S_{2k}$.

Let us start with the Gauss law constraints and try to resolve it
against the variable ${\bf{A}}_+ $.
But, because the vector
${\bf{A}}_- $ is the zero mode of these equations, only two
components of ${\bf{A}}_+$, transverse to ${\bf{A}}_- $, can be fixed uniquely.
Indeed, in our parametrization (\ref{eq:gpa1}), (\ref{eq:gpa2}),
when the ${\bf{A}}_- $ direction coincides with the ${\bf n}$ direction,
projection of the Gauss law equations by the transverse vectors $({\bf l}$ and ${\bf m})$
yields
\begin{eqnarray}
\label{eq:glc1}
&&
x \left[ ({\bf l \cdot h})-(\bf{l} \cdot {\bf{A}}_+) \right]=0\,,\\
\label{eq:glc2}
&&
x \left[ ({\bf m\cdot h})-(\bf{m} \cdot {\bf{A}}_+) \right]=0\,,
\end{eqnarray}
while its projection to the third ${\bf n}$-component results in constraint on the
variables $S_{ik}$
\begin{equation}\label{eq:algbc}
S^2_{11} + S^2_{12} + S^2_{21} + S^2_{22}=0\,.
\end{equation}
In (\ref{eq:glc1}) and (\ref{eq:glc2}) the helicity vector
${\bf{h}}=\bf{\dot{n}}\times\bf{n}$ has been introduced. So, from
(\ref{eq:glc1}) and (\ref{eq:glc2}), supposing $x\neq0$, it
follows that
\begin{equation}
{\bf{A}}_+= {\bf{h}}+ f {\bf{n}}\,,
\end{equation}
with unspecified function $f$.

With this ${\bf{A}}_+$ one can rewrite the equation of motion
(\ref{eq:E-Le}) for the ${\bf{A}}_-$ component as
({\it summation over $k=1,2$})
\begin{equation}\label{eq:egmx}
{\bf{n}}\,\ddot{x} = {\bf{A}}_k \times {\bf{\dot{A}}}_k-
{\bf{A}}_k({\bf{h}}\cdot{\bf{A}}_k)\,.
\end{equation}
Using (\ref{eq:E-Lcc}) the projection of the equation
(\ref{eq:egmx}) onto the ${\bf{n}}$ direction looks as
\begin{equation}\label{eq:prnx}
\ddot{x}=\frac{1}{x^2}\left(
({\bf{A}}_1\cdot{\bf{A}}_1)({\bf{A}}_2\cdot{\bf{A}}_2)-
({\bf{A}}_1\cdot{\bf{A}}_2)({\bf{A}}_2\cdot{\bf{A}}_1)\right)\,,
\end{equation}
while its projections onto ${\bf{l}}$ and ${\bf{m}}$ direction
give again the constraint (\ref{eq:algbc}).

Consider now the equations (\ref{eq:E-Le}), projection to ${\bf
l}$ and ${\bf m}$ directions results in equations ({\it no
summation $i\neq k$})
\begin{eqnarray}\label{eq:s21}
\dot{x}S_{2k}+2x\dot{S}_{2k}+2x\left[({\bf{l}}\cdot{\bf{\dot{m
}}})-f\right]S_{1k}&=&
S_{1k}(S^2_{1i}+S^2_{2i})-S_{1i}(S_{1i}S_{1k}+S_{2i}S_{2k})\,,\\
-\dot{x}S_{2k}-2x\dot{S}_{2k}+2x\left[({\bf{l}}\cdot{\bf{\dot{m
}}})-f\right]S_{2k}&=&
S_{2k}(S^2_{1i}+S^2_{2i})-S_{2i}(S_{1i}S_{1k}+S_{2i}S_{2k})\,,
\label{eq:s22}
\end{eqnarray}
while projection onto the third direction $\bf{n}$ gives
\begin{eqnarray}\label{eq:s23}
 ({\bf{n}} \cdot {\bf{\dot{m}}})S_{1k}-
({\bf{n}} \cdot {\bf{\dot{l}}})S_{2k}&=& ({\bf{h}} \cdot
{\bf{\dot{l}}})S_{1k}+ ({\bf{h}} \cdot {\bf{\dot{m}}})S_{2k}\,.
\end{eqnarray}
The last two equations (\ref{eq:s23}) are identically satisfied
when the relations
\begin{equation}
{\bf{h}} \cdot {\bf{\dot{l}}} = {\bf{n}} \cdot {\bf{\dot{m}}}\,, \qquad
{\bf{h}} \cdot {\bf{m}} = - {\bf{n}} \cdot {\bf{\dot{l}}}
\end{equation}
are taken into account. One can make the equations (\ref{eq:s21})
and (\ref{eq:s22}) independent of the frame $({\bf l}\,,{\bf m}\,,
{\bf{n}})$ if the function $f$ in the ${\bf{A}}_+$ decomposition
is chosen as
\begin{equation}
f = {\bf{l}} \cdot {\bf{\dot{m}}} + \rho\,,
\end{equation}
where the function $\rho$ is still unspecified.

With this identification the following system of differential
equations for the unknown functions
$S_{11}\,,S_{12}\,,S_{21}\,,S_{22}\,$ arises
\begin{eqnarray}\label{eq:1}
\dot{x} S_{11}+2x\dot{S}_{11}-2x\rho S_{21}&=&
-S_{11}(S_{11}S_{22}-S_{12}S_{21})\,, \\
\dot{x} S_{12}+2x\dot{S}_{12}-2x\rho S_{22}&=&\quad
S_{11}(S_{11}S_{22}-S_{12}S_{21})\,,
 \label{eq:ss22}\\
\dot{x} S_{21}+2x\dot{S}_{21}+2x\rho S_{11}&=&
-S_{22}(S_{11}S_{22}-S_{12}S_{21})\,, \\
\dot{x} S_{22}+2x\dot{S}_{22}+2x\rho S_{12}&=&\quad
S_{21}(S_{11}S_{22}-S_{12}S_{21})\,.\label{eq:4}
\end{eqnarray}
Introduction of the new functions $$S_{ij}\sqrt{x} = Y_{ij}$$
removes the derivatives of the function $x$
\begin{eqnarray}\label{eq:11}
\dot{Y}_{11}-\rho Y_{21}&=&\quad\frac{Y_{12}}{2 x^2}\,
(Y_{11}Y_{22}-Y_{12}Y_{21})\,, \\
\dot{Y}_{12}-\rho Y_{22}&=&-\frac{Y_{11}}{2x^2}\,
(Y_{11}Y_{22}-Y_{12}Y_{21})\,,
\label{eq:m1}\\
\dot{Y}_{21}+\rho Y_{11}&=&\quad\frac{Y_{22}}{2 x^2}\,
(Y_{11}Y_{22}-Y_{12}Y_{21})\,, \label{eq:m2}\\
\dot{Y}_{22}+\rho Y_{12}&=&-\frac{Y_{21}}{2 x^2}\,
(Y_{11}Y_{22}-Y_{12}Y_{21})\,.
\label{eq:41}
\end{eqnarray}
Now one can specify the function $\rho$.
Due to the symmetry condition (\ref{eq:symm}), $Y_{12}=Y_{21}$,
this leads to the relation
\begin{equation}
\rho = \frac{1}{2x^2}\, (Y_{11}Y_{22}-Y_{12}Y_{21})\,.
\end{equation}
Moreover, because the system of equations
(\ref{eq:11})-(\ref{eq:41}) possesses the following first integral
\begin{equation}\label{eq:fi}
Y_{11}Y_{22} - Y_{12}Y_{21} = \mu\,, \qquad \mu := \mbox{constant}\,,
\end{equation}
one can express $\rho$ solely in terms of the $x$ variable
\begin{equation}
\rho=\frac{\mu}{2x^2}\,.
\end{equation}
Therefore, finally the equations (\ref{eq:11})-(\ref{eq:41})
reduce to a system of three differential equations
\begin{eqnarray}\label{eq:111}
\dot{Y}_{11}&=&\frac{\mu}{x^2}Y_{12}\,, \\
\dot{Y}_{12}&=&\frac{\mu}{2x^2}\left(Y_{22}-Y_{11}\right)\,,\nonumber\\
\dot{Y}_{22}&=-&\frac{\mu}{x^2}Y_{12}\,.\nonumber
\end{eqnarray}
These equations should be solved together with the algebraic
constraint (\ref{eq:algbc}) which states
\begin{equation}
Y^2_{11}+ 2\,Y^2_{12} + Y^2_{22}=0\,.
\end{equation}
The last equation has only trivial {\em real} solutions
\begin{equation}
Y_{11} = Y_{12} = Y_{22} = 0
\end{equation}
that lead to the free equation of motion for the $x$ variable.

However, one can consider the analytic continuation of our
variables $Y$ in a complex domain. So, relaxing the reality
conditions, we use the following parametrization for the {\em
complex} $Y$-functions
\begin{eqnarray}\label{eq:fi1}
&& Y_{12}= \nu (i-1)\sin\chi\cos\chi\,, \qquad
Y_{11}= \nu(\cos^2\chi+ i \sin^2\chi)\,,\\
&& Y_{21}= \nu (i-1)\sin\chi\cos\chi\,,\qquad
Y_{22}=\nu(\sin^2\chi + i \cos^2\chi)\,. \label{eq:fi2}
\end{eqnarray}
Here the parameter $\nu$  is expressed through the first integral
constant  $\nu^2 =- i\mu$.

Within the parametrization (\ref{eq:fi1}) and (\ref{eq:fi2}) all
equations (\ref{eq:111}) are satisfied if the angular variable
$\chi$ obeys the equation
\begin{equation}\label{eq:chi}
\dot\chi= \frac{\mu}{2x^2} \,.
\end{equation}
Now in order to insure the self consistency of the solution  it
may be checked that the right hand side of the equation
(\ref{eq:prnx}) evaluated with $Y$ given in (\ref{eq:fi1}) reduces
to
\begin{equation}\label{eq:xmo}
\ddot{x}= \frac{\mu^2}{x^3}\,.
\end{equation}
This completes the proof, since (\ref{eq:xmo}) is the equation of
motion for an one-dimensional system with the  Lagrange function
\begin{equation}\label{eq:lcm}
L_{CM}:=\frac{1}{2}\,\bigg({\dot{x}}^2-\frac{\mu^2}{x^2}\bigg)\,.
\end{equation}


\end{document}